\documentclass[
reprint,
 amsmath,amssymb,
 aps,
 superscriptaddress,
prb,
revtex4
]{revtex4-2}

\usepackage{graphicx}
\usepackage{dcolumn}
\usepackage{bm}
\usepackage{physics}
\usepackage{xcolor}

\newcommand{\Rp}{\mathbf{R_p}}
\newcommand{\Rel}{\mathbf{R_e}}

\begin{document}

\title{Electron-Phonon Coupling in Correlated Metals: \\ A Dynamical Mean-Field Theory Study}

\author{David J. Abramovitch}%
\email{dabramovitch@caltech.edu}
\affiliation{Department of Applied Physics and Materials Science, California Institute of Technology, Pasadena, California 91125}
\affiliation{Center for Computational Quantum Physics, Flatiron Institute, 162 5th Avenue, New York, New York 10010, USA}
\author{Jennifer Coulter}
\affiliation{Center for Computational Quantum Physics, Flatiron Institute, 162 5th Avenue, New York, New York 10010, USA}
\author{Sophie Beck}
\affiliation{Center for Computational Quantum Physics, Flatiron Institute, 162 5th Avenue, New York, New York 10010, USA}
\affiliation{Institute of Solid State Physics, TU Wien, 1040 Vienna, Austria}
\author{Andrew Millis}
\email{amillis@flatironinstitute.org}
\affiliation{Center for Computational Quantum Physics, Flatiron Institute, 162 5th Avenue, New York, New York 10010, USA}
\affiliation{Department of Physics, Columbia University, New York, New York 10027, USA}

\date{\today}

\begin{abstract}
Strong electron-electron interactions are known to significantly modify the electron-phonon coupling relative to the predictions of density functional theory, but this effect is challenging to calculate with realistic theories of strongly correlated materials. Here we define and calculate a version of the EPC applicable beyond band theory by combining first principles density functional theory plus dynamical mean-field theory with finite difference phonon perturbations, presenting results for several representative phonon modes in two materials of interest. In the three-orbital correlated metal SrVO$_3$, we find that intra-V-$t_{2g}$-band correlation significantly increases the coupling of these electrons to a Jahn-Teller phonon mode that splits the degenerate orbital energies, while slightly reducing the coupling associated with a breathing phonon that couples to the charge on each V atom. In the infinite layer cuprate CaCuO$_2$, we find that local correlation within the $d_{x^2-y^2}$ orbital derived band has a modest effect on coupling of near-Fermi surface electrons to optical breathing modes. In both cases, the interaction correction to the electron-phonon coupling predicted by dynamical mean-field theory has a significant dependence on the electronic frequency, arising from a lattice-distortion dependence of the correlated electron dynamics, showing the inadequacy of the simple picture in which correlations change static local susceptibilities. We also show that the electron-phonon scattering and phonon lifetimes associated with these phonon modes are modified by the electronic correlation. Our findings shed light on the material- and mode-specific role of dynamical electronic correlation in electron-phonon coupling and highlight the importance of developing efficient computational methods for treating electron-phonon coupling in correlated materials.
\end{abstract}

\maketitle

\section{Introduction}

Electron-phonon ($e$-ph) interactions play an essential role in the physics of metals~\cite{grimvall_electron_1981, giustino_electron_phonon_2017}. A current focus is quantum materials with electronic properties controlled by strong correlations arising from electron-electron ($e$-$e$) interactions. However, there is growing evidence that $e$-ph interactions contribute in important ways to phenomena observed in these materials. For example, $e$-ph interactions have been argued to cause the ``kinks" seen in angle-resolved photoemission studies of high-$T_c$ cuprates~\cite{lanzara_evidence_2001, li_unmasking_2021}, phase transitions in rare earth nickelates~\cite{merritt_giant_2020,Georgescu22,li_twogap_2024, zhong_electronically_2024}, important contributions to transport and optical lifetimes in some correlated metals~\cite{millis_dynamic_1996,quijada_optical_1998, zhao_electrical_2000, abramovitch_respective_2024,coulter_mechanisms_2025}, and contributions to novel superconductivity~\cite{yin_correlation-enhanced_2013, li_electron-phonon_2019,xu_superconductivity_2020,mandal_strong_2014, gerber_femtosecond_2017, ding_correlation_2022, li_twogap_2024}. Experiment and theory indicate that the strength of the $e$-ph interactions may be strongly affected by electronic correlations~\cite{reznik_photoemission_2008, li_unmasking_2021, yin_correlation-enhanced_2013, li_electron-phonon_2019, mandal_strong_2014, gerber_femtosecond_2017, ding_correlation_2022, li_twogap_2024}.
\\
\indent 
The current standard for $e$-ph calculations in real materials obtains electron-phonon coupling (EPC) from the dependence of the Kohn-Sham potential in density functional theory (DFT) on atomic displacements. However, EPC is sensitive to the treatment of electronic correlations, and DFT is not always sufficient. For example, calculations using DFT with hybrid or advanced functionals~\cite{laflamme_electronphonon_2010, yin_correlation-enhanced_2013, komelj_electronphonon_2015, wang_accurate_2024}, ``+$U$" extensions~\cite{floris_phonons_2020, zhou_ab_2021, abramovitch_respective_2024,chang_first_2024}, and $GW$~\cite{attaccalite_doped_2010, faber_electronphonon_2011,antonius_manybody_2014, li_electron-phonon_2019,li_unmasking_2021,li_electron_2024} have found modified EPC compared to semi-local DFT in a range of materials. However, these methods are perturbative and/or employ static correlation, and as a result, do not necessarily describe basic electronic properties of strongly correlated quantum materials (such as the large dynamical mass enhancement owing to the frequency-dependent self-energy).
 \\
 \indent
 Methods applicable to strong electronic correlation, such as dynamical mean-field theory (DMFT)~\cite{georges_dmft_1996}, have been used to study the physics of electron-lattice coupling in model systems \cite{huang_electron_phonon_2003,kulic_influence_1994,deppeler_dynamical_2002,sangiovanni_electronphonon_2005, sangiovanni_electronphonon_2006, bauer_quantitative_2011,kim_interplay_2006,li_competing_2017,scazzola_competing_2023, moghadas_effective_2025, coulter_electronphonon_2025}. It is natural to ask whether the combination of density functional and dynamical mean-field theory (DFT+DMFT), which has proven a powerful and effective method of calculating electronic properties of correlated materials~\cite{kotliar_electronic_structure_2006,held_electronic_2007}, can be used to study EPC in correlated systems. First principles DFT+DMFT calculations have  been shown to strongly modify proxies for EPC, such as spectral shifts in response to atomic displacements in FeSe~\cite{mandal_strong_2014, ding_correlation_2022}, and the energetics of octahedral tilting in SrMoO$_3$~\cite{hampel_correlation_2021} and optical phonons in LaNiO$_2$~\cite{wang_hardening_2025}. However, the EPC has never been calculated directly in real materials with DFT+DMFT, and its dependence on the microscopic nature of the relevant electronic states and phonon modes has not been investigated.
\\
\indent
Here, we present a method to calculate the EPC for selected phonon modes using DFT+DMFT calculations with finite phonon perturbations obtained from first principles Density Functional Perturbtion Theory (DFPT)~\cite{baroni_phonons_2001} calculations. We use the method to investigate the effect of correlation on EPC and related properties for representative phonon modes in two correlated materials of interest: SrVO$_3$ (SVO), a multi-orbital correlated metal where the interesting physics arises from bands derived from the V t$_{2g}$ levels and phonons may couple to the orbital splitting, and CaCuO$_2$ (CCO), a member of the high-$T_c$ cuprate family where the correlation physics primarily resides in a single Cu $d_{x^2-y^2}$-derived band.
\\
\indent 
In SVO, our calculations show that local correlation significantly increases the coupling of t$_{2g}$ electrons to a Jahn-Teller optical phonon mode which couples to the orbital degree of freedom, while slightly decreasing the coupling associated with a breathing mode which couples to the total V-site charge density. The frequency dependence of the coupling is non-negligible. We use the computed EPC to calculate temperature-dependent scattering rates of electrons due to these phonon modes. In the infinite layer cuprate CaCuO$_2$ (CCO), our calculations show that correlation within the $d_{x^2-y^2}$ orbital has a moderate effect on the coupling of Fermi surface electrons to full- and half-breathing modes. However, we find that the correlation effect depends strongly on electronic frequency. We investigate this behavior at different interaction strengths and carrier concentrations, and calculate phonon linewidths.
\\
\indent 
The rest of this paper is organized as follows: In the next section (Sec.~\ref{sec:theory}), we present our theoretical method, based on finite difference DFT+DMFT calculations. 
Then we present the results of these calculations in SVO with the application to the temperature-dependent electron scattering (Sec.~\ref{sec:srvo3}), and in CCO investigating the filling and $U$ dependence and the phonon linewidth (Sec.~\ref{sec:cacuo2}). We next discuss in Sec.~\ref{sec:impurity} the interpretation of our method and results in terms of the auxiliary  impurity model used in the DFMT calculations  and then compare to DFPT+$U$ and other research on correlation-modified $e$-ph interactions in Sec.~\ref{sec:comparison}. Finally, we summarize our findings in Sec.~\ref{sec:conclusion}.

\section{Theory and Methods}\label{sec:theory}

\subsection{Definition of Coupling}

In this paper, as in most treatments of EPC in solids, we make the Born-Oppenheimer approximation, valid when the ratio of a typical ionic velocity to a typical electronic velocity is very small. In this case the EPC may be approximated by the response of the electronic system to {\it static} ionic displacements. In DFT-based EPC calculations the EPC is defined as the scattering amplitude between two Kohn-Sham states induced by a small atomic displacement. In a strongly interacting system, the electronic physics should instead be described in terms of a Green's function (propagator), and the EPC is correspondingly defined in terms of changes to the electron Green's function induced by changes in atomic position.
\\
\indent
The frequency-domain electron Green's function $\hat{G}(r',r;\omega)$ describing  the propagation of an electron from $r$ to $r^\prime$ is defined in terms of a reference non-interacting system, which in this paper we take to be the Kohn-Sham Hamiltonian $\hat{T}+\hat{V}_{KS}$ with $\hat{T}$ the free electron kinetic energy operator and $\hat{V}_{KS}$ a Kohn-Sham potential. The differences between the physical propagation and the propagation defined by the reference non-interacting system are parametrized by the self-energy operator $\hat{\Sigma}(r,r^\prime;\omega)$, thus 
\begin{equation}
G(r',r;\omega) = \left[\omega + \mu - \hat{T} - \hat{V}_{KS} - \hat{\Sigma}(\omega)\right]^{-1} 
\label{eq:Gdef}
\end{equation}
where $\mu$ is the chemical potential and the carat $\hat{} $ denotes an operator acting on spatial indices. A displacement $u_{\Rp a\mu}$ of atom $a$ in unit cell at $\Rp$ in direction $\mu$ changes both $\hat{V}_{KS}$ and $\hat{\Sigma}$. A formal expansion to first order in $u_{\Rp a\mu}$ (using the identity $\delta G = -G[\delta G^{-1}] G$) gives,
 \begin{multline}
    G(r',r;\omega) 
     =  G^{cn}(r',r;\omega) \\
     + \int dr^{\prime\prime}dr^{\prime\prime\prime} u_{\Rp a\mu} G^{cn}(r',r'';\omega) \times \\ \frac{\partial}{\partial u_{\Rp a\mu}} \left[\hat{V}_{KS} + \hat{\Sigma}(\omega) \right] G^{cn}(r''',r;\omega) + \mathcal{O}(u^2).
     \label{eq:G_rs_expansion}
\end{multline}
Here and below, we use $cn$ (for "clamped nuclei") to denote quantities in the absence of atomic displacements. The object sandwiched between the two unperturbed Green's functions, $\frac{\partial}{\partial u_{\Rp a\mu}} \left[\hat{V}_{KS} + \hat{\Sigma}(\omega) \right]$ is the many-body definition of the EPC. In the absence of many-body effects, the self-energy operator vanishes so that the Green's function describes propagation of Kohn-Sham eigenstates and the definition reduces to the standard DFT EPC.
\\
\indent 
In our calculations, we use a basis of DFT eigenstates, denoted $\ket{\phi_{n\mathbf{k}}}$ for band $n$ at electron crystal momentum $\mathbf{k}$, and express atomic displacements in terms of normal mode phonons with index $\nu$ and crystal momentum $\mathbf{q}$. Represented this way, our definition of the EPC $g^{\mathbf{k}\mathbf{q}}_{mn\nu}(\omega)$ in a correlated system is
 \begin{equation}
     g^{\mathbf{k}\mathbf{q}}_{mn\nu}(\omega) = \bra{\phi_{m\mathbf{k}+\mathbf{q}}} \partial_{\nu\mathbf{q}} \left[\hat{V}_{KS} + \hat{\Sigma}(\omega) \right] \ket{\phi_{n\mathbf{k}}}_{u.c.}
     \label{eq:coupling-def}
 \end{equation}
which now depends on the electronic frequency due to the frequency dependence of the Green's function and here the carat $\hat{}$ denotes an operator in the momentum indices. $\partial_{\nu\mathbf{q}}$ denotes the standard phonon mode-resolved derivative
  \begin{equation}
      \partial_{\nu\mathbf{q}} = \sum_{\Rp a\mu}e^{i\mathbf{q}\cdot \Rp} \sqrt{\frac{\hbar}{2M_a \omega_{\nu\mathbf{q}}}} \mathbf{e}^{a \mu}_{\nu \mathbf{q}} \left. \frac{\partial}{\partial u_{\Rp a\mu}}\right |_{u_{\Rp a\mu} = 0}
      \label{eq:displacementoperator}
  \end{equation}
 where $a$ indexes an atom with mass $M_a$, $\Rp$ denotes a unit cell, and $\omega_{\nu\mathbf{q}}$ and $\mathbf{e}^{a \mu}_{\nu \mathbf{q}}$ are respectively the phonon energy and normalized eigenvector.  Similar forms of EPC have been used for model systems~\cite{huang_electron_phonon_2003} and first-principles $GW$-based calculations~\cite{li_electron-phonon_2019}. 

\subsection{Evaluation with Dynamical Mean-Field Theory}
The central challenge  is to reliably evaluate the change in the self-energy due to the phonon perturbation $\partial_{\nu\boldsymbol{q}}\Sigma(\omega)$. 
We evaluate this using the density functional plus dynamical mean-field theory (DFT+DMFT) approximation. In DFT+DMFT, one combines a band structure calculated with DFT with a dynamical mean-field treatment of strong correlations within a set of local orbitals $\phi_i(r - \mathbf{\tau}_a -\Rel)$ that are constructed around an atom $a$ located at position $\mathbf{\tau}_a$ within unit cell $\Rel$ using Wannier or projector methods. The DMFT assumption is that the electronic self-energy is site-local in the orbital basis (but may be a matrix in the on-site orbital indices $ij$), so that the upfolded self-energy is given by
\begin{equation}
\Sigma(r,r^\prime;\omega) = \sum_{\Rel aij}\phi^{*}_i(r-\mathbf{\tau}_a-\Rel)\Sigma^{a}_{ij}(\omega)\phi_{j}(r-\mathbf{\tau}_a-\Rel).
\label{eq:SigmaDMFT}
\end{equation}
The self-energy $\Sigma^{a}_{ij}(\omega)$ for the correlated site is calculated from the solution of a quantum impurity model defined by the DFT band structure, the local orbitals, a  site-local inter-orbital  interaction representing the effective screened Coulomb interaction projected onto these orbitals, and a ``double-counting correction", each of which can be treated in several approximations. 
\\
\indent
In this paper we use the ``frontier orbital" DFT+DMFT methodology in which the correlated orbitals are defined by a Wannierization of bands crossing the Fermi level, we neglect full charge self-consistency, and we subtract a double-counting term from the self-energy. We use empirical interaction parameters and neglect changes to the interaction parameters associated with phonon displacements for simplicity. Similar approximations are commonly used in DFT+DMFT studies of structural deformations~\cite{mandal_strong_2014, kocer_efficient_2020, hampel_correlation_2021,Georgescu22}.
\\
\indent
The materials studied in this paper have a single correlated site (a transition metal) in the undistorted unit cell, with a small number of relevant correlated orbitals (partially filled $d$-shells). We consider phonon modes corresponding to lattice distortions which reduce the translational symmetry of the unit cell, resulting in a supercell with multiple correlated atoms. We apply the phonon perturbation of magnitude $\alpha$ to the supercell by adding $\alpha \times \bra{i,\Rel}\frac{1}{2}(\partial_{\nu\mathbf{q}}\hat{V}_{KS} + \partial_{\nu\mathbf{-q}}\hat{V}_{KS})\ket{j,\Rel'}$ to the supercell Wannier Hamiltonian. Here, $\alpha$ represents the ratio of the effective atomic displacements in the finite perturbation to the phonon eigen-displacements, and the term in the bra-ket is the Wannier basis perturbation caused by a set of real atomic displacements~\footnote{The phonon modes studied in this paper have $\mathbf{q} = -\mathbf{q}$ and thus the corresponding atomic displacements $u_{\Rp a\mu} = \sqrt{\frac{\hbar}{2M_a \omega_{\nu\mathbf{q}}}}e^{i\mathbf{q}\cdot \Rp} \mathbf{e}^{a \mu}_{\nu \mathbf{q}}$ can be written as fully real. However, for a general phonon, real atomic displacements $\sqrt{\frac{\hbar}{2M_a \omega_{\nu\mathbf{q}}}} \frac{1}{2}(e^{i\mathbf{q}\cdot \Rp} \mathbf{e}^{a \mu}_{\nu \mathbf{q}} + e^{-i\mathbf{q}\cdot \Rp} \mathbf{e}^{a \mu}_{\nu \mathbf{-q}})$ must be used to obtain a physical perturbed Hamiltonian. This is described as the real valued displacement field in Ref.~\cite{giustino_electronphonon_2007}.} associated with the phonon in question calculated with DFPT~\cite{baroni_phonons_2001} and Wannier interpolation~\cite{giustino_electronphonon_2007}.
\\
\indent
 We perform DFT+DMFT calculations with the perturbed and unperturbed Hamiltonians and extract the self-energies corresponding to the inequivalent sites in the perturbed supercell ($\hat{\Sigma}^{ph}(\omega)$) and unperturbed supercell ($\Sigma^{cn}(\omega)$). The self-energy term of the EPC in Eq.~\ref{eq:coupling-def} is then calculated from the difference between the two aforementioned sets of site-dependent self-energies, up-folded into the band basis as

 \begin{multline}
    \bra{\phi_{m\mathbf{k}+\mathbf{q}}} \partial_{\nu\mathbf{q}} \hat{\Sigma}(\omega)\ket{\phi_{n\mathbf{k}}}_{u.c.} = \frac{1}{\alpha N_e}\sum_{\Rel}e^{-i\mathbf{q}\cdot \Rel}  \\ \times U_{mi}^\dagger(\mathbf{k}+\mathbf{q}) \bra{i,\Rel} \hat{\Sigma}^{ph}(\omega) - \hat{\Sigma}^{cn}(\omega) \ket{j,\Rel}U_{jn}(\mathbf{k}) ,
    \label{eq:sigma-pert-fd}
\end{multline}

where $U(\mathbf{k})$ are the transformations between the Wannier and Bloch basis and $\alpha$ is chosen to be small enough to generate a good approximation to the derivative. Note that in this procedure the Wannier basis is held fixed.

Here we have used standard $e$-ph Fourier transform conventions~\cite{giustino_electronphonon_2007} but with the simplification that the DFT+DMFT self-energy is local and therefore non-zero only for $\Rel = \Rel'$. The DFT+DMFT EPC is then calculated with Eq.~\ref{eq:coupling-def} using Eq.~\ref{eq:sigma-pert-fd} and $\bra{\phi_{m\mathbf{k}+\mathbf{q}}} \partial_{\nu\mathbf{q}} \hat{V}_{KS} \ket{\phi_{n\mathbf{k}}}_{u.c.}$ from DFPT (which is interpolated from the Wannier function form in the standard way~\cite{giustino_electronphonon_2007}). The workflow for performing finite difference DFT+DMFT calculations is further explained in the Supplemental Material (SM)~\cite{supplemental_material}.

\subsection{Electron-Phonon Scattering}
The EPC calculated in our approach can be used to calculate quantities such as the electron and phonon scattering rates due to the $e$-ph interaction, which are related to the imaginary part of the $e$-ph contribution to the electron and phonon self-energies. To lowest order, the electron self-energy is a single phonon exchange, which can be schematically written $\Sigma = ig^2GD$, where $g$ is the EPC, $G$ is the electron Green's function, and $D$ is the phonon Green's function. When evaluated with the frequency-dependent DFT+DMFT EPC, DFT+DMFT electron Green's function, and the non-interacting DFPT phonon Green's function, the imaginary part is, 
  \begin{multline}
     \mathrm{Im}\Sigma^{e\mathrm{-ph}}_{\mathbf{k},nn'}(\omega) = \\\frac{-\pi}{N_\mathbf{q}}\sum_{\mathbf{q}\nu m} \Big[ g^{\mathbf{k}\mathbf{q}}_{mn\nu}\left(\omega-\frac{\omega_{\nu\mathbf{q}}}{2}\right)^* g^{\mathbf{k}\mathbf{q}}_{mn'\nu}\left(\omega-\frac{\omega_{\nu\mathbf{q}}}{2}\right) \\ \times A^{\mathrm{DMFT}}_{m\mathbf{k}+\mathbf{q}}(\omega - \omega_{\nu\mathbf{q}})[n(\omega_{\nu\mathbf{q}}) + 1 - f(\omega - \omega_{\nu\mathbf{q}})]  \\ + 
     g^{\mathbf{k}\mathbf{q}}_{mn\nu}\left(\omega+\frac{\omega_{\nu\mathbf{q}}}{2}\right)^* g^{\mathbf{k}\mathbf{q}}_{mn'\nu}\left(\omega+\frac{\omega_{\nu\mathbf{q}}}{2}\right) \\ \times  A^{\mathrm{DMFT}}_{m\mathbf{k}+\mathbf{q}}\left(\omega + \omega_{\nu\mathbf{q}}\right)\left[n(\omega_{\nu\mathbf{q}}) + f(\omega + \omega_{\nu\mathbf{q}})\right] \Big].
     \label{eq:elph-se-reomega}
 \end{multline}
where $A(\omega)$ is the spectral function and $f$ and $n$ are respectively the fermion and boson occupation numbers. Eq.~\ref{eq:elph-se-reomega} reduces to the conventional DFT Fan-Migdal expression if $A(\omega)$ is taken to be a Dirac delta function at the band energy and the  EPC is defined from $\hat{V}_{KS}$ only~\cite{migdal_interaction_1958, giustino_electron_phonon_2017}. Previous work also studied $e$-ph interactions in correlated materials using the full DMFT spectral function but the DFPT EPC~\cite{abramovitch_combining_2023}. 
\\
\indent
Similarly, to lowest order in $g$ the $e$-ph contribution to the phonon linewidth is given by a bubble diagram formed from two $e$-ph vertices and two electron propagators (schematically, $\Pi = g^2G^2$) calculated as,
 \begin{multline}
  \mathrm{Im}\Pi^{e\mathrm{-ph}}_{\nu\mathbf{q}}\left(\omega_{\nu\mathbf{q}}\right) = \frac{-2\pi}{N_\mathbf{k}}\sum_{mn\mathbf{k}}\int d\omega \left|g^{\mathbf{k}\mathbf{q}}_{mn\nu}\left(\omega + \frac{\omega_{\nu\mathbf{q}}}{2}\right)\right|^2  \\ \times A^{\mathrm{DMFT}}_{n\mathbf{k}}(\omega)A^{\mathrm{DMFT}}_{m\mathbf{k}+\mathbf{q}}(\omega + \omega_{\nu\mathbf{q}})\left[f(\omega) - f(\omega + \omega_{\nu\mathbf{q}})\right], 
    \label{eq:phonon-se}
\end{multline}

If $A(\omega)$ is set to be a Dirac delta function at the band energy and the DFPT EPC is used, Eq.~\ref{eq:phonon-se} is the conventional expression for the leading order electronic contribution to the phonon linewidth~\cite{allen_neutron_1972, giustino_electron_phonon_2017}. The importance of using a beyond-DFT (non-delta function) $A$ was noted in a recent work (which however used the DFPT-derived EPC)~\cite{park_nonadiabatic_2024}.
\\
\indent
In Eqs.~\ref{eq:elph-se-reomega} and~\ref{eq:phonon-se}, the EPC depends on one frequency because we have neglected the dependence on the phonon frequency due to the Born-Oppenheimer approximation. Since the coupling connects two electronic states that have different frequencies, we evaluate it at the average of the two electronic frequencies (i.e. $\omega$ and $\omega \pm \omega_{\nu\mathbf{q}}$) to avoid biasing the calculation. Because the two frequencies are separated by $\omega_{\nu\mathbf{q}}$, this approximation disappears in the $\omega_{\nu\mathbf{q}} \rightarrow 0$ limit. In our calculations, we calculate scattering rates using EPC derived from the real part of $\partial_{\nu\mathbf{q}}\Sigma(\omega)$ because imaginary part is small near the Fermi surface.
\\
\indent 
We also note that these forms of self-energy 
have two vertex corrected $e$-ph interactions as they come from perturbative expansions in phonon displacements. This differs from the self-energy in the Hedin-Baym equations, due to the latter's self-consistency~\cite{hedin_effects_1970, giustino_electron_phonon_2017}. A similar approximation using two screened vertices has been found to be accurate when considering electronic screening of $e$-ph interactions~\cite{berges_phonon_2023} and shown to be physically correct for the phonon lifetime~\cite{deppeler_dynamical_2002,stefanucci_exact_2025}.
\\
\indent
 The derivation and computational evaluation of these self-energies is further discussed in the Supplemental Material (SM)~\cite{supplemental_material}, as is the effect of the frequency averaging approximation, the use of the DFT vs. DFT+DMFT electronic Green's functions, and other approximations.

\subsection{Computational Methods \label{sec:computational-methods}}
We calculate the electronic structure, phonons, and EPC using DFT and DFPT with the {\sc Quantum Espresso} code~\cite{Giannozzi_Quantum_2009}.  We convert the EPC to the real-space Wannier basis and interpolate using {\sc Perturbo}~\cite{zhou_perturbo_2021}.  We use Wannier90~\cite{arash_updated_Wannier90_2014} to construct Wannier orbitals, which are used as the DMFT impurity orbitals. More details on the parameters used for DFT and DFPT are provided in the SM~\cite{supplemental_material}\nocite{garrity_pseudopotentials_2014}. 
\\
\indent
All DMFT calculations use TRIQS~\cite{parcollet_triqs_2015} with the TRIQS/DFTTools and solid$\_$dmft first-principles packages~\cite{aichhorn_triqsdfttools_2016, Merkel_solid_dmft_2022}. The lattice Hamiltonian is calculated by interpolating the Wannier Hamiltonian and summing over a sufficiently large $\mathbf{k}-$grid to converge the hybridization. All calculations use the continuous-time quantum Monte Carlo$-$hybridization expansion solver (CTHYB)~\cite{gull_ctmc_2011, seth_triqscthyb_2016}, the Held formula for the double counting correction~\cite{held_electronic_2007}, and Pad\'e analytic continuation. In the finite difference DFT+DMFT calculations, we use a scale factor of $\alpha = 1/2$ for the phonon perturbation for all modes. More details on the DFT+DMFT calculations are provided in the SM~\cite{supplemental_material}. 
\\
\indent
For SVO we use the experimental lattice parameter of $a = 3.842$~\AA~\cite{lan_structure_2003,ahn_low_energy_2022} in the $Pm\bar{3}m$ cubic structure. We compute maximally-localized Wannier functions for the three $t_{2g}$-derived orbitals. For the DMFT calculations, we use a Hubbard-Kanamori interaction Hamiltonian with $U = 4.5$ eV and $J = 0.675$ eV. Our DFT, DFPT, and unperturbed DFT+DMFT calculations in SVO largely follow previous work~\cite{abramovitch_respective_2024}. For the finite difference DFT+DMFT calculations, we use a 2 $\times$ 2 $\times$ 2 supercell for both modes.
\\
\indent 
For CCO, we use the tetragonal $P4/mmm$ structure with a hole doping fraction of 0.15 holes per stoichiometric unit cell and relaxed lattice parameters of $a \times a \times c$ = 3.742 \AA $\times$ 3.742 \AA $\times$ 3.051 \AA. We use an impurity consisting of the $d_{x^2-y^2}$ Wannier orbital and a Hubbard interaction. As described below, we perform calculations with a variety of Hubbard $U$ parameters (3.1, 3.9, and 4.7 eV), and consider different fillings in the rigid band approximation. For the finite difference DFT+DMFT calculations, we use a 2 $\times$ 2 $\times$ 1 supercell for both modes. 

\section{Results: $\mathbf{SrVO_3}$ \label{sec:srvo3}}

\subsection{Modes and Coupling}

\begin{figure}[h]
    \centering
    \includegraphics[width=\linewidth]{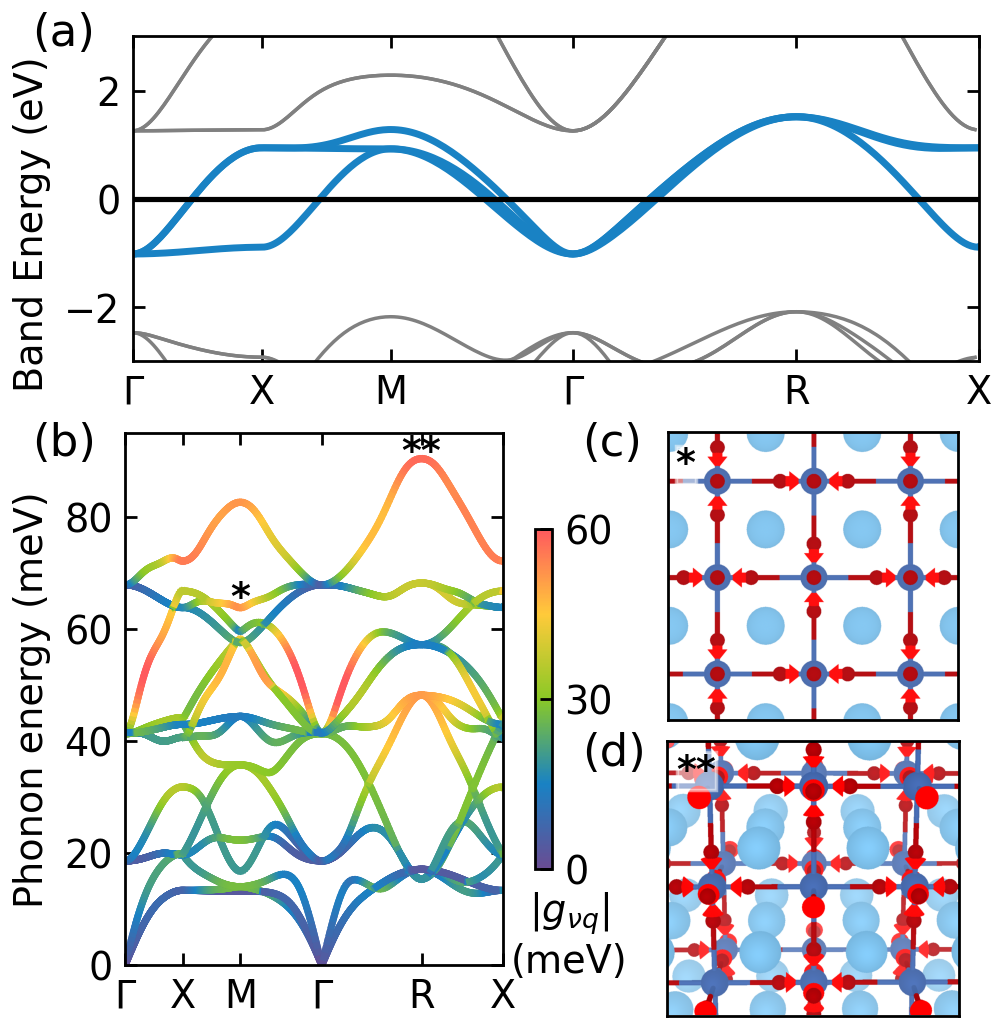}
    \caption{(a) Band structure of SrVO$_3$, highlighting in blue the Wannierized t$_{2g}$ bands used in the DMFT calculations. (b) Phonon dispersion of SrVO$_3$, with color showing EPC averaged on the Fermi surface as in \cite{abramovitch_respective_2024}. 
    The $M$ and $R$-point phonon modes studied are indicated are indicated by single and double stars, respectively.  Visualizations of (c) the $\mathbf{M}$-point bond stretching mode and (d) the $\mathbf{R}$-point breathing mode (right) where red atoms are O, light blue are Sr, and dark blue are V. }
    \label{fig:svo_bands_eph_phvis}
\end{figure}

SVO is a correlated metal with a cubic perovskite structure and one electron in the V-$t_{2g}$ orbitals  often used as a prototypical example for many-body methods. SVO stands out as a promising system to study the interplay of $e$-$e$ and $e$-ph interactions, as demonstrated by recent calculations showing $e$-ph-induced photoemission kinks and phonon-limited resistivity above $\sim$25 K~\cite{abramovitch_respective_2024} and experiments showing transport properties beyond Fermi liquid $e$-$e$ scattering~\cite{brahlek_hidden_2024}. In the present context, the multi-orbital electronic structure of the t$_{2g}$ orbitals (respective bands shown in Fig.~\ref{fig:svo_bands_eph_phvis}(a)) is of interest because it enables investigation of both ``Jahn-Teller"-type phonons that lift the orbital degeneracy and ``Holstein"-type phonons that couple mainly to the on-site charge.
\\

\indent
We focus on two representative phonon modes in SVO: an $\mathbf{M}$-point Jahn-Teller mode which couples to the orbital degree of freedom on each site, and an $\mathbf{R}$-point breathing mode which couples to the charge degree of freedom on alternating sites. These modes are indicated on the DFPT phonon dispersion in Fig.~\ref{fig:svo_bands_eph_phvis}(b), where color indicates the EPC (averaged over electronic states on the Fermi surface as in Ref.~\cite{abramovitch_respective_2024}). Both modes are high-frequency ($\omega_\mathbf{M} = 64$ meV and $\omega_\mathbf{R} = 90$ meV) optical phonon modes with fairly strong EPC, owing to stretching V$-$O bonds. The atomic displacements associated with each mode are illustrated in Fig.~\ref{fig:svo_bands_eph_phvis}(c,d). The pattern of O displacements in the $\mathbf{M}$-point Jahn-Teller mode (c) splits the three t$_{2g}$ orbitals on each V site while maintaining equivalence between all sites (under the combination of translation and $\pi/2$ rotation), whereas in the $\mathbf{R}$-point breathing mode (d), alternating V sites are split but the orbitals on each site remain equivalent. 
\\
\indent
\begin{figure*}
    \centering
    \includegraphics[width=\textwidth]{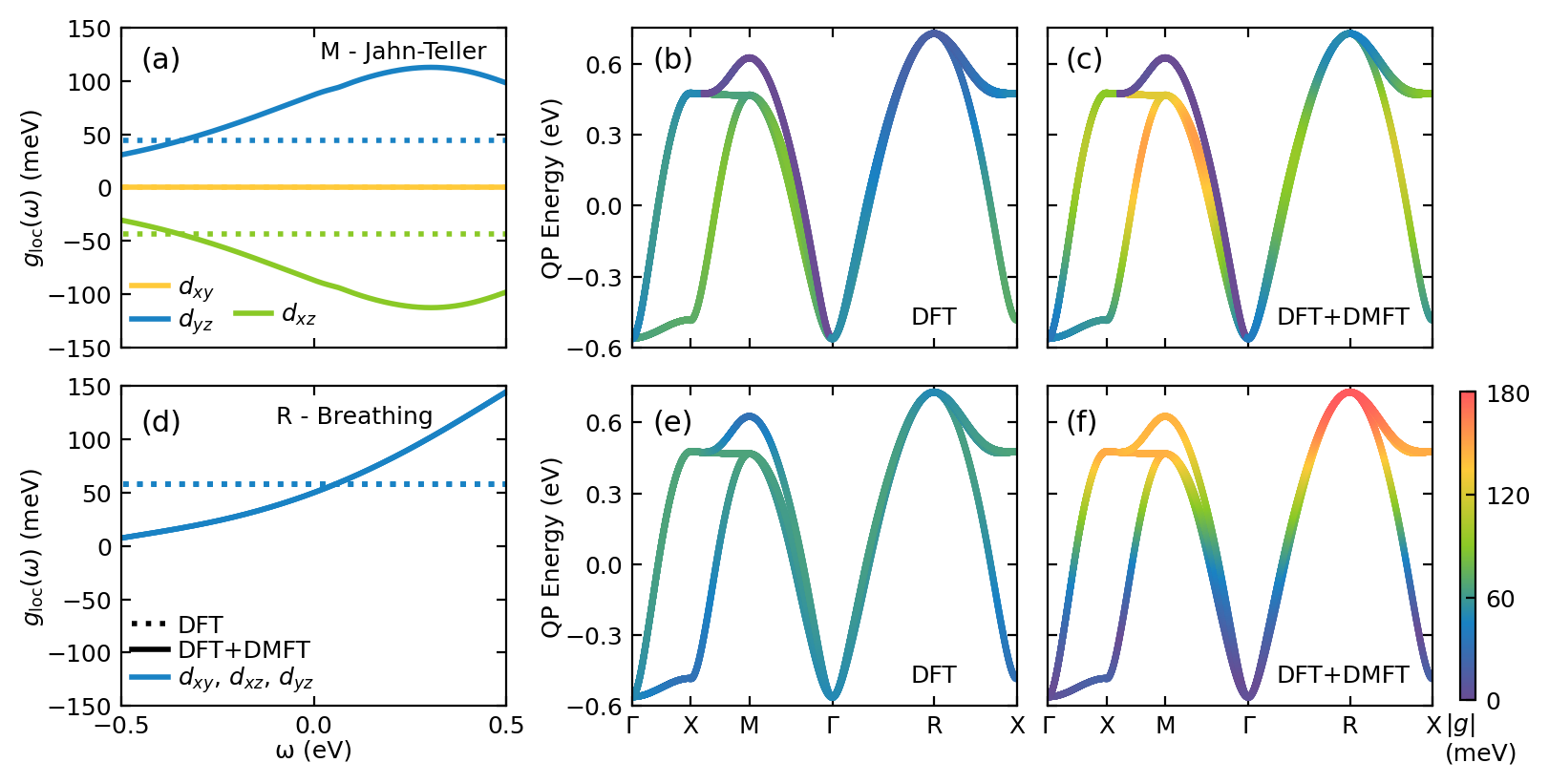}
    \caption{EPC in SVO with DFT and DFT+DMFT. (a) Local EPC in the Wannier basis for the $\mathbf{M}$-point mode calculated with DFT (dashed lines, $\partial_\mathbf{\nu\mathbf{q}}V(\mathbf{R} = 0)$) and DFT+DMFT (solid lines, $\partial_\mathbf{\nu\mathbf{q}} \left[ V(\mathbf{R} = 0) + \Sigma(\omega,\mathbf{R} = 0)\right]$). (b) DFT and (c) DFT+DMFT EPC projected onto the quasi-particle bands $|g| = (\sum_m |g^{\mathbf{k}\mathbf{q}}_{nm\nu}|^2)^{1/2}$. (d) Local EPC and band-resolved (e) DFT and (f) DFT+DMFT EPC for the $\mathbf{R}$-point breathing mode. Calculations shown are for $\beta$ = 80 / eV. 
    }
    \label{fig:svo_epc_dft_dmft}
\end{figure*}
We perform finite difference DFT+DMFT EPC calculations for both modes and compare to DFPT. Fig~\ref{fig:svo_epc_dft_dmft}(a) shows the Wannier basis local EPC $\bra{i,0} \partial_{\nu\mathbf{q}}[V_{KS} + \Sigma(\omega)]\ket{j,0}$ $-$ where $i,j$ label orbital indices and the site is $\Rel = 0$; this quantity provides a natural comparison between DFT and DFT+DMFT because the DMFT contribution is local in the Wannier basis. For the $\mathbf{M}$-point Jahn-Teller mode, we find significantly increased EPC to electrons near the Fermi surface. As expected, the mode splits the d$_{yz}$ and d$_{xz}$ orbitals while leaving the $d_{xy}$ orbital unchanged.  The Wannier basis local coupling strength is nearly doubled from DFT (44 meV) to DFT+DMFT at $\omega = 0$ (87 mev). The coupling also exhibits a fairly strong dependence on the electron frequency, increasing in strength with increasing frequency. The band-resolved EPC calculated with DFT and DFT+DMFT is plotted on the quasi-particle bands (defined as the solutions of $\det[\omega + \mu - \varepsilon_{n\mathbf{k}}\delta_{nn'} - \mathrm{Re}\Sigma^{\mathrm{DMFT}}_{nn'\mathbf{k}}(\omega)] = 0$) in Fig.~\ref{fig:svo_epc_dft_dmft}(b) and (c), respectively. The DFT EPC shows clear band dependence arising from the orbital character of the bands and non-local terms in the coupling. This dependence is maintained in the DFT+DMFT EPC, but there is a clear enhancement of the coupling strength as well as frequency dependence from the self-energy term.
\\
\indent
In contrast to the $\mathbf{M}$-point mode, the $\mathbf{R}$-point breathing mode shows a slight decrease in the magnitude of the local EPC, from 58 meV with DFT to 50 meV with DFT+DMFT at $\omega=0$ eV (Fig.~\ref{fig:svo_epc_dft_dmft}(d)). Like the $\mathbf{M}$-point mode, the coupling shows a frequency dependence which tends to increase coupling at higher frequencies, but exhibits less band and $\mathbf{k}$ dependence due to the symmetry of the displacements. The band-resolved DFT and DFT+DMFT EPC are plotted on the quasi-particle bands in Fig.~\ref{fig:svo_epc_dft_dmft}(e) and (f), respectively.
\\
\indent 
In summary, DMFT significantly increases the coupling of the $\mathbf{M}$-mode which couples to orbital splitting, but slightly decreases the coupling of the $\mathbf{R}$-mode which couples to charge disproportionation, and introduces a strong frequency dependence to the coupling, absent in DFPT or its $+U$ and hybrid functional extensions. The difference in correlation effects on the EPC for the two modes may be understood in context of local interactions enhancing orbital fluctuations and suppressing charge fluctuations in correlated systems. The strong frequency dependence is another important consequence of correlation physics and will be discussed in more detail below.

\subsection{Electron Scattering}

\begin{figure}
    \centering
    \includegraphics[width=\linewidth]{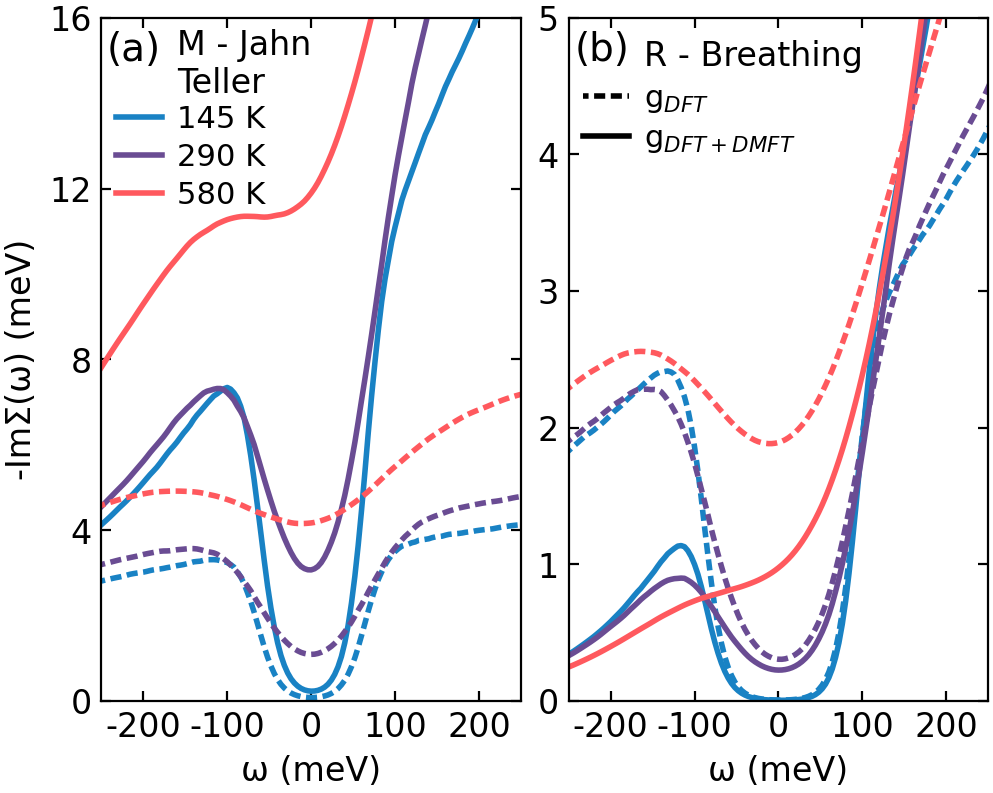}
    \caption{Contribution to the electron scattering rates in SVO calculated using the DFT+DMFT couplings (solid lines) and DFPT coupling (dashed lines) from (a) scattering by the $\mathbf{M}$-point bond stretching mode and (b) $\mathbf{R}$-point breathing mode. The frequency-dependent scattering rates calculated using DFT and DFT+DMFT EPC are shown as the BZ averaged $-\mathrm{Im}\Sigma(\omega)$ in the Wannier orbital basis for the $d_{xz/yz}$ orbital for the $\mathbf{M}$-point mode and the $d_{xy/xz/yz}$ orbital for the $\mathbf{R}$-point mode.}
    \label{fig:svo_elph_se}
\end{figure}
Next, we calculate the scattering of electrons by each of the two phonon modes studied. The scattering is proportional to the imaginary part of the lowest order self-energy (Eq.~\ref{eq:elph-se-reomega}); 
we calculate the contributions from the phonon modes studied using the EPC calculated above.  
 We plot the Brillouin Zone (BZ) averaged part of this self-energy in the Wannier basis $\Sigma_{ii}(\omega) = \sum_\mathbf{k} U_{in}(\mathbf{k}) \Sigma_{nn'}(\mathbf{k},\omega) U_{n'i}(\mathbf{k})^\dagger$ in Fig.~\ref{fig:svo_elph_se} at several temperatures (the temperature applies to both the DFT+DMFT calculation and the electron and phonon occupation numbers in Eq.~\ref{eq:elph-se-reomega}). 
\\
\indent
We find that the electron scattering near the Fermi energy by the $\mathbf{M}$ mode phonon (Fig.~\ref{fig:svo_elph_se}(a)) is significantly increased when calculated with DFT+DMFT EPC compared to DFT EPC, while scattering by the $\mathbf{R}$ mode phonon (Fig.~\ref{fig:svo_elph_se}(b)) is slightly suppressed, consistent with our calculated coupling strengths. We also find a significantly increased asymmetry between scattering above and below the Fermi energy due to the frequency dependence of the EPC, with the scattering rate at positive frequencies enhanced, consistent with the frequency dependence of the EPC. For both modes, the scattering rate obeys a temperature dependence consistent with the electron and phonon occupation numbers, with the temperature dependence of the DFT+DMFT EPC having a minor effect (See SM~\cite{supplemental_material}). We note that these calculations used the DFPT phonon frequency; the phonon frequency enters the results both via the convention for $g$ (Eq.~\ref{eq:displacementoperator}) and via the phonon propagator. From these formulae the effects of substituting the physical phonon frequencies may be deduced. Based on a previous study~\cite{kocer_efficient_2020}, correlation results in a moderate hardening of the optical phonons in SVO, which could suppress scattering independent of the coupling used.

\section{Results: $\mathbf{CaCuO_2}$ \label{sec:cacuo2}}

\subsection{Modes and Coupling}
Next, we investigate the EPC in the DFT+DMFT theory of calcium copper oxide (CCO), a member of the  widely studied family of cuprate superconductors. In cuprate materials, $e$-$e$ interactions are believed to dominate the physics. These interactions have been extensively studied with DMFT,  both in the context of minimal models with a single $d_{x^2-y^2}$ orbital per Cu site~\cite{grilli_meanfield_1990,lichtenstein_antiferromagnetism_2000, capriotti_effect_2005,Seneschal05, Haule07,Sordi12,Gull13} and more advanced calculations incorporating additional orbitals, dynamically screened interactions, and/or $GW$~\cite{Emery88,Weber10,Wang11,choi_first_2016}. The effect of $e$-ph interactions in cuprates have been studied in relation to superconductivity~\cite{song_electronphonon_1995, savrasov_linear_1996}, quasi-particle properties~\cite{yam_dressing_2022,chang_first_2024} including as a potential origin of photoemission kinks~\cite{lanzara_evidence_2001, giustino_small_2008,reznik_photoemission_2008, li_unmasking_2021}, and phonon softening associated with metal-insulator transition or charge density waves~\cite{mcqueeney_anomalous_1999, pintschovius_anomalous_1999, falter_effect_2000, zhang_electron_phonon_2007, reznik_giant_2010, mansart_temperature_2013}. The cuprates thus provide a useful computational platform to study many-body effects on the EPC in a single-orbital strong correlation context.
\\
\indent 
CCO is the structurally simplest cuprate-based superconductor. The basic  unit is a flat Cu-O$_2$ plane with a $C_4$ symmetry and a single Cu ion per unit cell. The interplane coupling is very weak so that the $d_{x^2-y^2}$ electronic band is nearly two-dimensional~\cite{azuma_superconductivity_1992,karpinski_single_1994}. CCO is insulating in the undoped normal state, but upon hole doping transitions from insulating to metallic, where a range of doping values in the metallic phase also show superconducting behavior.
 \\
 \indent
 \begin{figure}
     \centering
     \includegraphics[width=\linewidth]{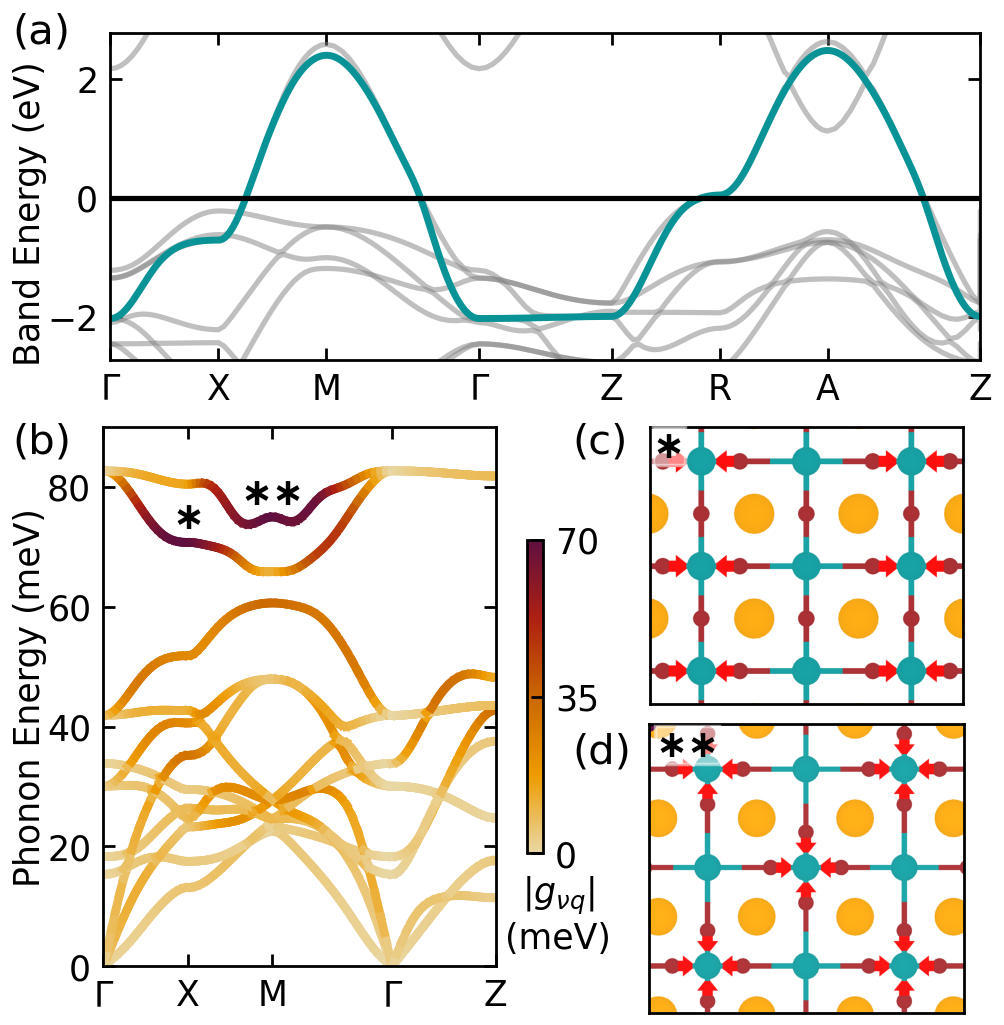}
     \caption{(a) Band structure of CaCuO$_2$, showing the Wannierized d$_{x^2-y^2}$ band used in the DMFT calculations (teal) and other bands in the DFT calculation (grey). (b) DFPT-calculated phonon dispersion of CaCuO$_2$, with color showing EPC averaged on the Fermi surface and studied half breathing (*) and full breathing (**) phonon modes indicated. Visualizations of (c) the $\mathbf{X}$-point half breathing mode and (d) the $\mathbf{M}$-point full breathing mode. Cu atoms are shown in teal, O in red, and Ca in orange.}
     \label{fig:cco_bands_eph_phvis}
 \end{figure}
 We base our DMFT calculations on a minimal model of the electronic structure comprised of the $d_{x^2-y^2}$ Wannier orbital, which dominates the electronic structure near the Fermi energy as shown in Fig.~\ref{fig:cco_bands_eph_phvis}(a). We calculate the phonon dispersion and the EPC with DFPT, as shown in Fig.~\ref{fig:cco_bands_eph_phvis}(b). The two high-symmetry optical phonon modes which couple most strongly to the $d_{x^2-y^2}$ band at the Fermi surface are the $\mathbf{X}$-point half-breathing mode (panel (c)), and an $\mathbf{M}$-point full-breathing mode (panel (d)) marked by single and double stars in panel (b), respectively. Both modes involve movement of O atoms in the Cu-O plane as depicted in Fig.~\ref{fig:cco_bands_eph_phvis}(c) and (d). We perform finite difference DFT+DMFT EPC calculations on these modes. 
 \\
 \indent
 \subsection{$U$-dependent EPC at 0.15 hole doping}
\begin{figure*}
     \centering
     \includegraphics[width=1.0\textwidth]{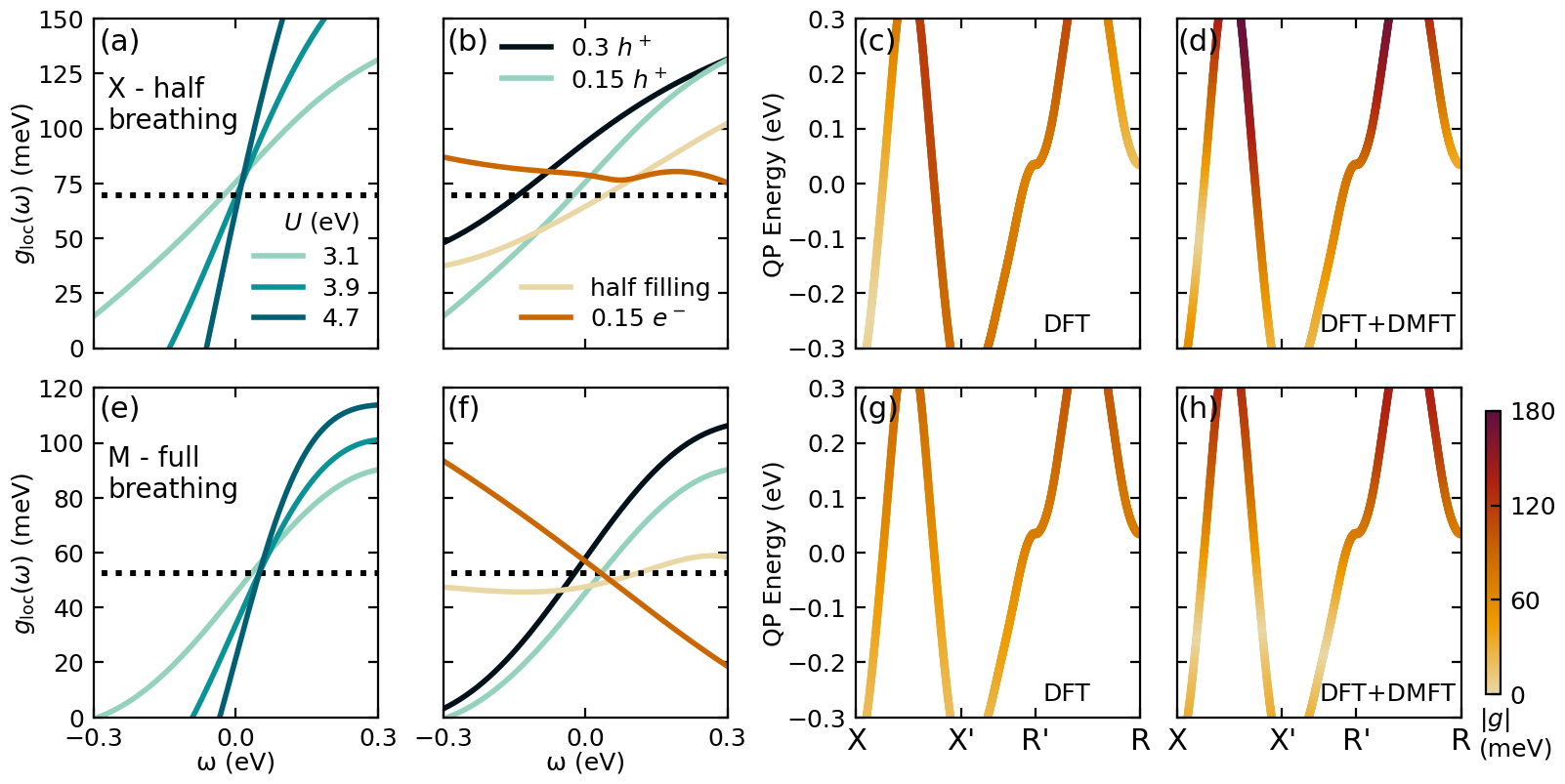}
      \caption{EPC in CCO with DFT and DFT+DMFT. (a) Local EPC in the Wannier basis for the $\mathbf{X}$-point half breathing mode calculated with DFT (dashed lines, $\partial_\mathbf{\nu\mathbf{q}}V(\mathbf{R} = 0)$) and DFT+DMFT (solid lines, $\partial_\mathbf{\nu\mathbf{q}} \left[ V(\mathbf{R} = 0) + \Sigma(\omega,\mathbf{R} = 0)\right]$) as a function of electron frequency $\omega$ at $U$-values indicated at 0.15 hope doping and (b) at the filling values indicated and $U$ = 3.1 eV. (c) DFT and (d) DFT+DMFT EPC at 0.15 hole doping and $U$ = 3.1 eV projected onto the quasi-particle bands $|g| = (\sum_m |g^{\mathbf{k}\mathbf{q}}_{nm\nu}|^2)^{1/2}$ (e-h). Same as (a-d) but for the $\mathbf{M}$-point full breathing mode.}
     \label{fig:cco_epc_dft_dmft}
 \end{figure*}

 Fig.~\ref{fig:cco_epc_dft_dmft} shows the EPC at several $U$ values for $0.15$ hole doping, a carrier concentration at which high T$_c$ superconductivity occurs in some of the cuprate compounds.  We begin by considering the Wannier basis local EPC  for $U = 3.1 $ eV, shown in light blue in Fig.~\ref{fig:cco_epc_dft_dmft}(a) for the $\mathbf{X}$-point half-breathing mode and Fig.~\ref{fig:cco_epc_dft_dmft}(e) for the $\mathbf{M}$-point full-breathing mode. For both phonons we observe that the difference in the EPC at the Fermi level between DFT and DFT+DMFT is small, but that the DMFT-derived EPC has an extremely strong frequency dependence over a few hundred meV scale. The highly frequency-dependent EPC combines with the $\mathbf{k}$ dependence already present in DFPT and yields a strong frequency dependence when projected onto the quasiparticle bands shown in panels (c,d) ($\mathbf{X}$-point mode) and (g,h) ($\mathbf{M}$-point mode). 
 \\
 \indent
We now consider  higher $U$-values. The Wannier local EPC at $U$ = 3.1, 3.9, and 4.7 eV is shown in Fig.~\ref{fig:cco_epc_dft_dmft}(a) for the $\mathbf{X}$-point half-breathing mode and Fig.~\ref{fig:cco_epc_dft_dmft}(e) for the $\mathbf{M}$-point full-breathing mode. While the zero frequency EPC shows moderate changes compared to  the DFT value for all values of U, the frequency dependence increases sharply with increasing $U$. This frequency dependence becomes so strong that for $U$ = 4.7 eV, the $g_{\mathrm{loc}}(\omega)$ varies from 0 to $\sim2 \times g_{loc}(\omega = 0)$ in a range of $\pm \omega_{\nu\mathbf{q}} \sim 70 \mathrm{~meV}$ from the Fermi energy. This indicates the potential for large adiabaticity in the $e$-ph interaction at large $U$, i.e. because the $e$-ph interaction varies significantly on the scale of the phonon frequency, the approximation of a static phonon for the purposes of calculating coupling may be insufficient. 
\\
\indent
The increasing frequency dependence of $g$ can be connected with increasing correlation strength; $Z$ decreases from $\sim 0.50$ at $U$ = 3.1 eV to $\sim 0.28$ at $U$ = 4.7 eV corresponding to an increased slope in the real part of the self-energy (see SM~\cite{supplemental_material}). The increasing frequency dependence indicates that the coupling of the phonon to correlation increases with the degree of correlation. We note that our calculations show only a moderate change in the $\omega = 0$ coupling, although the change might be larger in more advanced calculations including more orbitals or dynamical screening$-$this is discussed in Sec.~\ref{sec:comparison}. 
 
\subsection{Filling-dependent EPC at $U = 3.1$ eV}

The dependence of the Wannier-local  EPC on carrier concentration is shown for $U=3.1$ eV in panels (b,f) of Fig.~\ref{fig:cco_epc_dft_dmft}. The zero frequency coupling shows some variation as a function of filling, with generally larger DFT+DMFT couplings further away from half-filling. The striking behavior is that  for both phonon modes the frequency dependence has a strong dependence on carrier concentration, with a large, positive slope at $0.3$ and $0.15$ hole-doping, then flattening as the carrier concentration moves across half filling and then changing sign for increasing electron doping.    
\\
\indent 
The observed interplay of the carrier concentration and the magnitude and sign of the frequency dependence suggests a close link between the frequency dependence of $g$ and the frequency dependence of the self-energy. For a Fermi liquid, the dominant feature in the low frequency electron self-energy is a linear frequency dependence, Re$\Sigma(\omega)=-(Z^{-1}-1)\omega$, with (in the DMFT approximation) the coefficient $Z < 1$ giving the inverse mass enhancement and the quasiparticle weight and $Z^{-1}$ defining an effective correlation strength. In the metallic state of a Mott-Hubbard-like system, the correlation strength increases and $Z$ decreases with increasing $U$. For $U$ values near the Mott transition critical $U$, correlation is strongest (and $Z$ is smallest) near half-filling and decreases with either electron or hole doping. This trend is evident in our calculated $\mathrm{Re}\Sigma(\omega)$ in CCO (see SM~\cite{supplemental_material}). Because the  phonon mode couples to the site occupancy, it is natural to suppose that the main contribution to $\partial_{\nu\mathbf{q}}\Sigma(\omega)$ is via the dependence of $Z$ on orbital occupation $n$, suggesting a contribution to $g\sim \frac{\partial Z^{-1}\omega}{\partial n } \frac{\partial n}{\partial u_{\nu\mathbf{q}}}$. The derivative increases with increasing $Z^{-1}$ (and hence with $U$), is small near half-filling where $Z$ passes through a minimum, and has a sign that depends on the sign of the doping.
\\
\indent
This explanation may also apply to SVO, in which the t$_{2g}$ orbitals are at $1/6$ filling and there is significant particle hole asymmetry.
\subsection{Phonon Linewidth}

\begin{figure}
    \includegraphics[width = \linewidth]{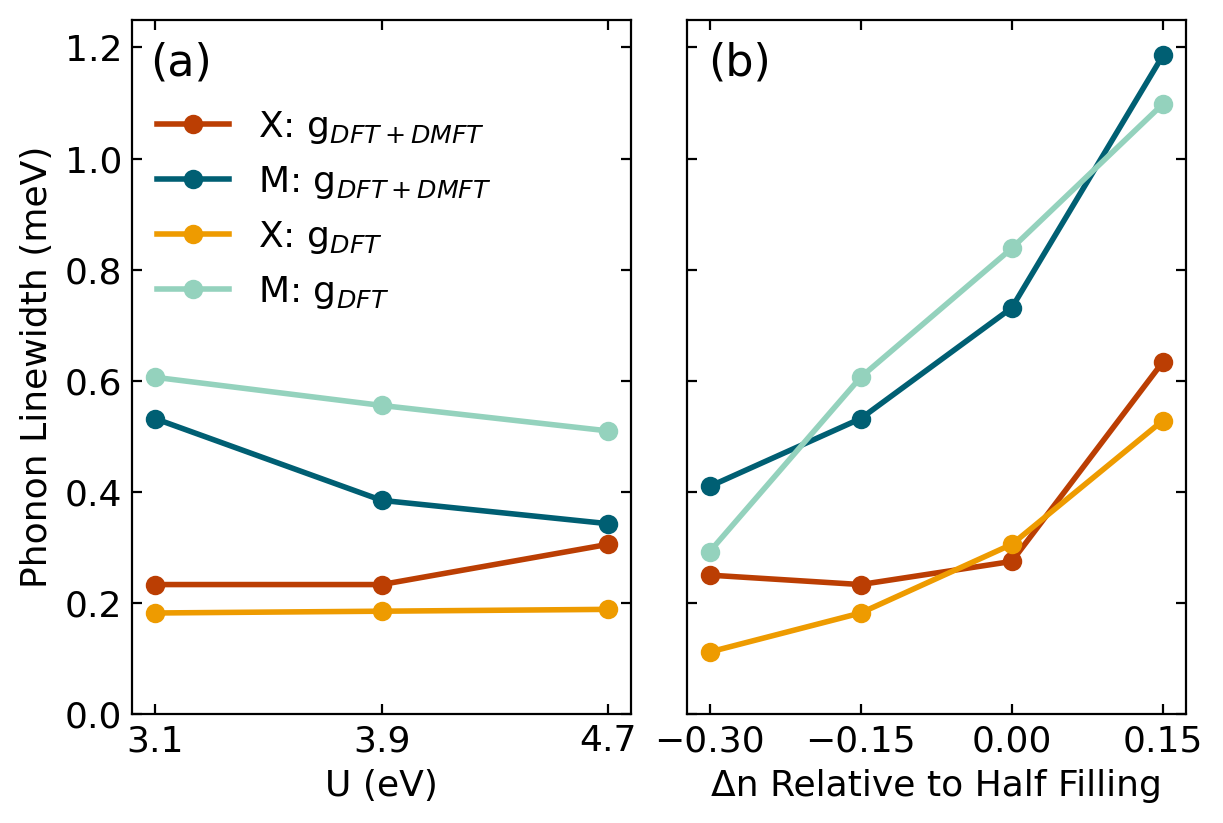}
    \caption{Phonon linewidth in CCO  (a) as a function of $U$ at 0.15 hole doping and (b) as a function of filling at $U$ = 3.1 eV.}
    \label{fig:cco-phonon-lw}
\end{figure}

As an application of the EPC in CCO, we calculate the $U$- and filling-dependent phonon linewidths $\gamma_{\nu\mathbf{q}}$ for the phonons studied. We evaluate $\gamma_{\nu\mathbf{q}} = 2 \mathrm{Im}\Pi_{\nu\mathbf{q}}(\omega_{\nu\mathbf{q}})$ with the phonon self-energy from Eq.~\ref{eq:phonon-se} based on the lowest order phonon self-energy diagram. The resulting linewidths are plotted in Fig.~\ref{fig:cco-phonon-lw}(a) as a function of $U$ and Fig.~\ref{fig:cco-phonon-lw}(b) as a function of filling.
\\
\indent
The dependence of the linewidth on correlation ($U$) is opposite for the two phonon modes. We see that for the $\mathbf{M}$-mode (dark blue), increasing correlation suppresses the linewidth at fixed doping. However, for the $\mathbf{X}$-mode (dark orange) an increase with $U$ is found. Here the linewidth reflects that the correlation increases the coupling of the $\mathbf{X}$-mode slightly and decreases that of the $\mathbf{M}$-mode slightly.
\\
\indent 
The linewidth of both phonons increases significantly with filling for the DFT EPC (light blue and orange lines) due to the increasing joint density of states at the wavevectors $\mathbf{q}$ as the Fermi energy is raised. When the DFT+DMFT EPC is used (dark blue and red lines), the trend remains but the linewidths increase or decrease slightly depending on the change to $g$ near the Fermi energy. 
Experimentally, an anomalously large phonon linewidth has been observed in La$_{1.85}$Sr$_{0.15}$CuO$_4$ at a $\mathbf{q}$-point between $\mathbf{\Gamma}$ and $\mathbf{X}$~\cite{reznik_giant_2010}, which was not found in DFT based calculations~\cite{reznik_photoemission_2008}. Future studies with DFT+DMFT may be able shed light on this observation, which appears to be related to strong correlation. 

\section{Electron-Phonon Coupling in the Impurity Problem \label{sec:impurity}}
While the main focus of this paper is on the correlation-induced changes in the lattice EPC, in this section we consider the EPC from the perspective of the quantum impurity model that determines the self-energy within DMFT. We first analyze the hybridization between the impurity and lattice $\Delta(i\omega_n)$ and its phonon-induced change $\partial_{\nu\mathbf{q}}\Delta(i\omega_n)$ in order to understand how lattice EPC translates to a coupling to the correlated local orbitals. Then, we analyze the self-energy $\Sigma(i\omega_n)$ and its phonon-induced change $\partial_{\nu\mathbf{q}}\Sigma(i\omega_n)$ to understand the interaction contribution to EPC in terms of changes to dynamical correlation. We work with the Matsubara-frequency quantities that are direct outputs from our calculations and draw mainly qualitative conclusions. 
 \begin{figure*}[bt]
    \centering
    \includegraphics[width=1.0\linewidth]{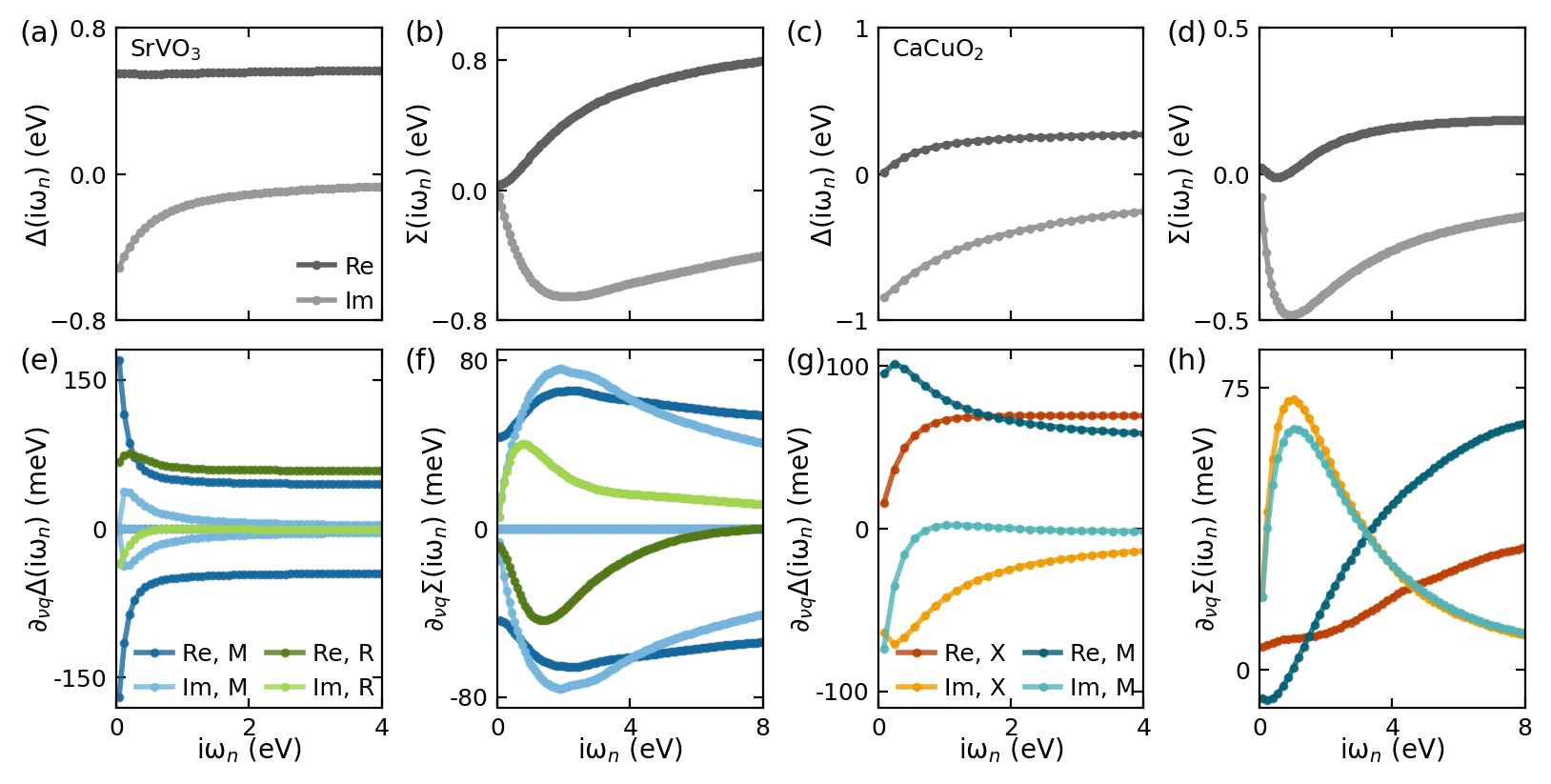}
    \caption{Impurity problem self-energy and hybridization on the Matsubara axis. (a) Impurity hybridization and (b) impurity self-energy in SVO at $\beta = 80$/ eV. (c-d) Same for CCO at 0.15 hole doping and $U = 3.1$ eV. (e) phonon perturbations to the impurity hybridization and (f) impurity self-energy in SVO. (g-h) Same for CCO. In all cases, the on-site energy shift is added into the hybridization, and the double counting is subtracted from the self-energy.}
    \label{fig:impurity}
\end{figure*}
 \\
 \indent
 In the impurity model, the local orbitals couple to a non-interacting bath of lattice states via a hybridization function $\Delta(i\omega_n)$. The hybridization is a frequency-dependent matrix in the local orbital basis determined through a self-consistency condition between the impurity and lattice Green's functions. A phonon-induced change to the non-interacting lattice states $\partial_{\nu\mathbf{q}}V_{\mathrm{KS}}$ will therefore translate to a change to the hybridization $\partial_{\nu\mathbf{q}}\Delta(i\omega_n)$. The change in the hybridization function defines the EPC in the impurity model. The infinite frequency limit of $\partial_{\nu\mathbf{q}}\Delta(i\omega_n)$ is a real matrix whose diagonal elements are phonon induced changes in the on-site energies~\footnote{In some formulations of DMFT the hybridization function is defined to vanish in the infinite frequency limit and the infinite frequency limit is included in the static Hamiltonian of the impurity model. In the present discussion we consider $V_{\mathrm{loc}}$ to be part of the hybridization as both it and the frequency-dependent part of $\Delta(i\omega_n)$ contribute an EPC in the impurity model.}, equal to the Wannier-local part of the DFPT EPC $\bra{i,0} \partial_{\nu\mathbf{q}}V_{KS} \ket{j,0}$. The frequency dependence of the change in $\Delta(i\omega_n)$ found in the our calculations reflects changes to the hybridization with surrounding sites due to a phonon distortion with a definite wavevector. These changes arise from phonon-induced changes to the intersite hopping, to in on-site energies on different sites, and to phonon-induced changes to correlation effects. The frequency dependence indicates that the physics of EPC to a specific $\mathbf{q}$-dependent phonon in the DFT+DMFT formalism is in general richer than models of local EPC.
 \\
 \indent 
Fig.~\ref{fig:impurity} panels (a) and (e) respectively show the Matsubara-axis hybridization function for SVO and the changes due to $\mathbf{M}$-mode and $\mathbf{R}$-mode phonon distortions. For the SVO $\mathbf{M}$-mode, which couples to level splitting between local orbitals, panel (e) shows that impurity EPC $\partial_{\nu\mathbf{q}}\Delta(i\omega_n)$ is primarily real. We observe that $\mathrm{Re}(\partial_{\nu q}\Delta(i\omega_n))$ is frequency dependent only below $\sim 0.5$ eV. The high frequency constant value is the DFPT Jahn-Teller coupling that relates atomic motions to the level splitting; the increase in amplitude at low frequency expresses the physics that band structure and correlation effects enhance the tendency towards orbital ordering at the $\mathbf{q} = \mathbf{M}$ wavevector.  For the SVO $\mathbf{R}$-mode, which couples to the total on-site charge density, $\partial_{\nu\mathbf{q}}\Delta(i\omega_n)$ is almost frequency-independent, suggesting that the charge susceptibility in SVO has only a weak momentum dependence.
\\
\indent 
Panels (c) and (g) of Fig.~\ref{fig:impurity} respectively show the Matsubara-axis hybridization function for CCO and the changes due to $\mathbf{X}$ and $\mathbf{M}$-mode phonon distortions. Panel (g) shows that for the CCO $\mathbf{M}$-mode (which does not break the point symmetry) the bare coupling is enhanced at low frequencies, indicating that lattice effects and non-local EPC increases the tendency to charge ordering at the $\mathbf{M}$-wavevector. For the $\mathbf{X}$-mode (which reduces point symmetry from $C_4$ to $C_2$), the Matsubara frequency dependence is opposite, indicating that these effects suppress the ordering tendency at this wavevector. In CCO, $\partial_{\nu\mathbf{q}}\Delta(i\omega_n)$ is moderately renormalized from its bare DFT value by correlation via the self-consistency condition (see SM~\cite{supplemental_material}). 
\\
\indent
We now consider the impurity self-energy $\Sigma(i\omega_n)$ and its phonon-induced change $\partial_{\nu\mathbf{q}}\Sigma(i\omega_n)$. The self-energy is the sum of a real, frequency-independent contribution (the Hartree ($\Sigma_H$) and double counting ($\Sigma_{DC}$) terms) and a frequency-dependent part that originates from the beyond-Hartree many-body dynamics and vanishes in the infinite frequency limit. The correlation contribution to the EPC is defined in terms of the changes to the self-energy; in what follows we analyze our results to obtain insights into the nature of changes. 
\\
\indent
It is useful to consider the spectral representation linking the imaginary part of the real frequency self-energy $\mathrm{Im}\Sigma(\omega)$ to the Matsubara axis self-energies,

\begin{eqnarray}
\mathrm{Im}\Sigma(i\omega_n)&=&-i\omega_n\int \frac{d\omega}{\pi}\frac{\mathrm{Im}\Sigma(\omega)}{\omega_n^2+\omega^2}
\label{eq:KKImSigma}
\\
\mathrm{Re}\Sigma(i\omega_n)&=&\int \frac{d\omega}{\pi}\frac{\omega\mathrm{Im}\Sigma(\omega)}{\omega_n^2+\omega^2}+\Sigma_H-\Sigma_{DC}
\label{eq:KKReSigma}
\end{eqnarray}
Note that the $i\omega_n$ dependence in $\mathrm{Re}\Sigma(i\omega_n)$ requires particle-hole asymmetry: if $\mathrm{Im}\Sigma(\omega)=\mathrm{Im}\Sigma(-\omega)$, $\mathrm{Re}\Sigma(i\omega_n)$ in Eq.~\ref{eq:KKReSigma} is frequency independent.
\\
\indent 
Eqs.~\ref{eq:KKImSigma},\ref{eq:KKReSigma} show that changes to $\Sigma(i\omega_n)$ can arise from both changes to the Hartree/DC terms and from changes to $\mathrm{Im}\Sigma(\omega)$. Frequency-independent beyond-DFT methods including DFT+$U$ and hybrid functionals give frequency-independent EPC renormalizations related to those arising here from changes to $\Sigma_{H}-\Sigma_{DC}$. These changes may be thought of as coming from correlation-induced changes in average local susceptibilities (for example suppression of on-site charge fluctuations by a repulsive $U$ or correlation-induced enhancements of orbital splitting relating to correlation-induced orbital order).
\\
\indent
Lattice distortion-induced changes to $\mathrm{Im}\Sigma(\omega)$ lead to frequency dependence of the EPC. To understand the nature of these changes, it is useful to review the basic properties of $\mathrm{Im}\Sigma(\omega)$, which may roughly be characterized by an over-all amplitude, a frequency scale, and a particle-hole asymmetry. For simple moderately correlated metals at low temperature, we expect $\mathrm{Im}\Sigma(\omega)$ to vanish at zero frequency, to grow $\sim \omega^2$ at low frequency, pass through maxima at positive and negative frequency near $\omega\sim \pm \omega_0$, and then to decay to zero at high frequency. In this picture, the particle-hole asymmetry is encoded in different weights at the peaks at $\pm \omega_0$ and the overall scale of the self-energy comes from the peak amplitudes. A simple model that encodes this physics is, 
\begin{equation}
    \mathrm{Im}\Sigma(\omega) = (A+dA)\delta(\omega - \omega_0) + (A-dA)\delta(\omega + \omega_0) 
\end{equation}
where $A$ is an amplitude with dimension of energy squared, $\omega_0$ is a characteristic  frequency scale, and $dA$ parametrizes the particle-hole asymmetry. Particle-hole asymmetry may also manifest as a difference in the absolute value of the peak positions for positive and negative frequency; as will be seen, introducing different frequencies is not necessary for our analysis. The parameters can be fit from the Matsubara axis data in Fig~\ref{fig:impurity}, with $\omega_0$ linked to the frequency scale and $A$ and $dA$ setting the magnitudes of $\mathrm{Im}\Sigma(i\omega_n)$ and $\mathrm{Re}\Sigma(i\omega_n)$ respectively. A rough measure of the effective low-frequency correlation strength is the low-energy mass enhancement/quasiparticle renormalization $Z^{-1}=1-\frac{\partial \mathrm{Im}\Sigma}{\partial i \omega}\sim \frac{A}{\omega_0^2}$, so that vanishing $\omega_0$ causes the effective low-frequency correlation strength to diverge. Though simplistic, this model qualitatively accounts for the calculated frequency dependence of $\partial_{\nu\mathbf{q}}\Sigma(i\omega_n)$ and its simple shape is conceptually confirmed by preliminary data from real frequency solvers to be published elsewhere. 
\\
\indent 
For SVO, the self-energy of the undistorted structure, shown in panel (b) of Fig.~\ref{fig:impurity} is consistent with this simple picture, with an $\mathrm{Im}\Sigma$ characterized by a single frequency scale $\omega_0\sim 1.8$ eV and a high degree of particle-hole asymmetry (since the change in real part of $\Sigma(i\omega_n)$ from high to low frequency $\int \frac{d\omega}{\pi}\frac{\omega\mathrm{Im}\Sigma}{\omega^2}\sim \frac{dA}{\omega_0}$ is very similar in magnitude to the maximum value of $\mathrm{Im}\Sigma(i\omega_n=\omega_0)=\int \frac{d\omega}{\pi}\frac{\mathrm{Im}\Sigma\omega_0}{\omega_0^2+\omega^2}\sim \frac{A}{\omega_0}$). 
\\
\indent
Now consider $\partial_{\nu\mathbf{q}}\Sigma(i\omega_n)$ for SVO shown in Fig.~\ref{fig:impurity}(f), beginning with the $\mathbf{M}$-mode. For this mode, there is no change in charge density on the impurity so $\partial_{\nu\mathbf{q}}\Sigma_{\mathrm{DC}} = 0$. We see that $\partial_{\nu\mathbf{q}}\mathrm{Re}\Sigma(i\omega_n)$  has only a weak $i\omega_n$ dependence, meaning that the phonon-induced change in the self-energy is nearly particle-hole symmetric. The $\partial_{\nu\mathbf{q}}\Sigma_{H}$ is large, about the same size as the static part of $\partial_{\nu q}\Delta$, and acts to amplify the effect of the lattice distortion as expected from the Hartree-level enhancement of the orbital susceptibility. However, the $\partial_{\nu\mathbf{q}}\mathrm{Im}\Sigma(i\omega_n)$ is of comparable magnitude to the Hartree shift, and has a frequency dependence very similar to that of $\mathrm{Im}\Sigma(i\omega_n)$, indicating that $\partial_{\nu\mathbf{q}}\Sigma(i\omega_n)$ is in effect an increase in $A$ plus a similarly sized Hartree shift. 
\\
\indent
The effect for the SVO $\mathbf{R}$-mode distortion is different. Because this phonon couples to the total charge density, the double counting term is affected. The real part of the change in the Matsubara axis self-energy essentially vanishes at high frequency, meaning that $\partial_{\nu\mathbf{q}}\Sigma_{DC}$ and $ \partial_{\nu\mathbf{q}}\Sigma_{H}$ cancel out. The change in the real part also nearly vanishes at low frequency, so the change in $\mathrm{Im}\Sigma$ must be composed of two counterbalancing effects which add in the imaginary part and cancel in the real part: a change in the particle-hole asymmetry and a change in the characteristic frequency. That $\partial_{\nu q}\mathrm{Im}\Sigma(i\omega_n)$ decays much more rapidly in $i\omega_n$ than does $\mathrm{Im}\Sigma$ also shows that a change in frequency $\omega_0$ is an important effect. These qualitative considerations are further supported by the more detailed modeling of the self-energy and its changes presented in the SM~\cite{supplemental_material}.
\\
\indent
For CCO, the self-energy of the undistorted structure, shown in Fig.~\ref{fig:impurity}(d) is also mainly consistent with this simple picture. The weak low frequency structure evident in $\mathrm{Re}\Sigma(i\omega_n)$ indicates a small amplitude low-frequency contribution to the self energy, which we do not consider further here. The main feature is a $\mathrm{Im}\Sigma$ characterized by a single frequency scale $\omega_0\sim 1$ eV and a lower degree of particle-hole asymmetry than in SVO (since the change in $\mathrm{Re}\Sigma(i\omega_n)$ from high to low frequency $\int \frac{d\omega}{\pi}\frac{\omega\mathrm{Im}\Sigma}{\omega^2}\sim \frac{dA}{\omega_0}$ is about half of the maximum value of $\mathrm{Im}\Sigma(i\omega_n=\omega_0)=\int \frac{d\omega}{\pi}\frac{\mathrm{Im}\Sigma\omega_0}{\omega_0^2+\omega^2}\sim \frac{A}{\omega_0}$).
\\
\indent 
Last we consider $\partial_{\nu\mathbf{q}}\Sigma(i\omega_n)$ for CCO shown in Fig.~\ref{fig:impurity}(h), beginning with the $\mathbf{M}$-mode. The large amplitude of the change in the real part, and the slower frequency decay (cf the discussion of the $\mathbf{R}$-mode in SVO) indicates that an important effect of the phonon distortion is a change in particle-hole asymmetry.  The more rapid decrease of the change in the imaginary part indicates that a particle-hole symmetric change in the basic frequency scale is also important. The $\mathbf{X}$-mode exhibits a smaller but still present change in particle hole asymmetry as well as a change in the frequency scale. Consistent with the discussion in Sec.~\ref{sec:cacuo2}, we see that the phonons couple to dynamical correlation including particle hole asymmetry and proximity to a Mott transition. 
\\
\indent 
This qualitative analysis of the self-energy reveals that the phonon-distortion-induced changes to the self-energy (and therefore the correlation-induced changes to the EPC) have a dynamical structure that depends on the physics of the material and the nature of the distortion. Because the phonon distortions can modify the correlation strength, particle hole asymmetry, and characteristic frequency of the self-energy in addition to inducing a static shift, a static (Hartree-type) theory of EPC misses important physical effects.

\section{Comparison to DFPT+$U$ and Previous Studies on Correlation-Modified EPC \label{sec:comparison}}

In this section we compare our results to DFPT+$U$ and other previous work, beginning with SVO. Our calculations show increased EPC and $e$-ph scattering for the $\mathbf{M}$-point Jahn-Teller phonon mode which couples to the orbital degree of freedom, while a slight suppression is found for the $\mathbf{R}$-point breathing mode. These effects can be qualitatively compared to those found in SVO using DFPT with a Hubbard correction~\cite{abramovitch_respective_2024} (differences in the interactions, orbitals, and charge self-consistency between the two calculations mean that the comparison is only qualitative). The Wannier-local EPC for these modes calculated with DFPT, DFT+DMFT, and DFPT+$U$ are shown in Table~I.
\begin{table}[h!]
\centering
 \begin{tabular}{||c c c||} 
 \hline
 $g_{\mathrm{loc}}$ & $\mathbf{M}$ Jahn-Teller   &  $\mathbf{R}$ Breathing  \\ [0.5ex] 
 \hline\hline
 DFPT & 44 meV & 58 meV \\ 
 DFPT+$U = 3$ eV~\cite{abramovitch_respective_2024, footnote_on_dfptU} & 73 meV & 61 meV  \\
 DMFT ($\omega = 0$)  & 87 meV & 50 meV \\
  DMFT ($\omega \rightarrow \infty$) &  93 meV & 60 meV \\
 \hline
 \end{tabular}
 \label{tab:SVO-gloc}
 \caption{Wannier-local EPC in SVO in different approximations. $g_{ij\nu}^{\Rel = 0, \mathbf{q}}$ is a diagonal matrix of (0, $g$, $-g$) for the $\mathbf{M}$-point mode and ($g$, $g$, $g$) for the $\mathbf{R}$-point mode, where $g$ is the value shown. DFT+DMFT calculations shown are at $\beta = 80$ / eV. DFPT+$U$ calculations use atomic orbitals with $U = 3$ eV and $J = 0$~\cite{abramovitch_respective_2024}.}
\end{table}
\\
\indent
DFPT+$U$ predicts some trends consistent with our DFT+DMFT calculations, such as significantly increased EPC for the $\mathbf{M}$-point mode compared to DFPT. However, DFPT+$U$ does not capture the frequency dependence of the coupling. One consequence of this can be seen in the $\mathbf{R}$-point mode: DFPT+$U$ and the $\omega \rightarrow \infty$ Hartree limit of DFT+DMFT both predict slight increases in the coupling strength, but DFT+DMFT predicts a slight decrease at $\omega = 0$ (which is more relevant to transport). This indicates that the frequency dependence is important because the EPC of physical interest near the Fermi energy may differ from that in the static limit. 
\\
\indent
In CCO, we find that $e$-$e$ interactions have a moderate effect on EPC at the Fermi energy and induce a strong frequency dependence. Some of these results are consistent with previous work in model systems, for example, our finding that increasing $U$ decreases the $\omega = 0$ EPC for the Holstein-like $\mathbf{M}$-point full breathing mode agrees with work in a Hubbard-Holstein model using quantum Monte Carlo~\cite{huang_electron_phonon_2003} and DMFT~\cite{coulter_electronphonon_2025}. Table~II lists the Wannier-local EPC in CCO in different approximations, including a DFPT+$U$ calculation. In DFPT+$U$, the coupling is slightly increased for both phonon modes compared to DFPT, which  can be compared with the $\omega \rightarrow \infty$ Hartree limit of DFT+DMFT where we find large increases for both modes. In contrast, at $\omega = 0$ in DFT+DMFT, the $\mathbf{X}$-mode coupling is slightly increased while the $\mathbf{M}$-mode coupling is slightly decreased, further demonstrating the importance of a dynamical treatment of the EPC.
\\
\begin{table}[h!]
\centering
 \begin{tabular}{||c c c||} 
 \hline
 $g_{\mathrm{loc}}$ & $\mathbf{X}$ Half-Breathing   &  $\mathbf{M}$ Full-Breathing  \\ [0.5ex] 
 \hline\hline
 DFPT &  70 meV &  53 meV \\ 
 DFPT+$U$~\cite{footnote_on_dfptU} &  79 meV &  65 meV  \\
 DMFT ($\omega = 0$)  &  76 meV &  45 meV \\
  DMFT ($\omega \rightarrow \infty$) & 117  meV & 137 meV \\
 \hline
 \end{tabular}
 \label{tab:CCO-gloc}
 \caption{Wannier local EPC in CCO in different approximations. All calculations have $U = 3.1$ eV and 0.15 hole doping. The DFPT+$U$ calculations use atomic orbitals.}
\end{table}
\indent
These results also highlight the effect of the double counting term in the self-energy. The Hartree contribution to the self-energy from the Hubbard interaction suppresses local charge fluctuations and thus decreases the effect of ``breathing mode" lattice distortions on the electrons, as observed in model-system studies~\cite{huang_electron_phonon_2003, coulter_electronphonon_2025}. However, the double counting term present in DFT+DMFT substantially changes the infinite frequency limit of the self-energy, partially canceling or in some cases outweighing the Hartree term, consistent with findings that the double counting correction in DFT+DMFT can promote charge ordering~\cite{carta_emergence_2024}. Further explanation of the double counting correction as well as tables indicating the contribution of the double counting potential to the Wannier-local $g$, and values of the Wannier-local $g$ for other temperatures in SVO and $U$ and doping values in CCO are provided in SM~\cite{supplemental_material}. In cases where the Wannierization contains both correlated and uncorrelated orbitals, the double counting corrections and charge self-consistency may play a more important role, making these important topics for future investigation.
\\
\indent
Recently, first-principles calculations using $GW$-based perturbation theory (GWPT) have found significantly increased EPC in hole doped La$_2$CuO$_4$~\cite{li_unmasking_2021} compared to DFPT, and both GWPT and hybrid functional calculations have found significantly increased EPC in single orbital bismuthates~\cite{yin_correlation-enhanced_2013, li_electron-phonon_2019}. These results, which have been attributed to decreased electronic screening in $GW$ and hybrid functional DFT compared to semi-local DFT, contrast our findings in CCO. While we expect our DFT+DMFT calculations to better capture local, dynamical correlation, they are limited by the single orbital impurity approximation and the DFT starting point. More research is needed to disentangle these effects on the $e$-ph interactions, and more realistic impurity construction schemes or methods like $GW$+DMFT~\cite{boehnke_strong_2016, zhu_abinitio_2021} may help clarify this by including both strong correlation and accurate electronic screening. 

\section{Conclusion \label{sec:conclusion}}
We have presented a method for calculating the EPC in a correlated material using DFPT and finite difference DFT+DMFT calculations. We applied this method to several representative optical phonon modes in two materials of interest: the 3-band cubic perovskite SVO, for which we compared phonons that couple to orbital splitting and to on-site density, and the infinite layer cuprate CCO, a representative Mott-Hubbard system.
\\
\indent
In SVO, we find that electronic correlation from DMFT significantly enhances the EPC at the Fermi level of an $\mathbf{M}$-point bond stretching phonon which couples to the orbital degree of freedom, while slightly suppressing that of an $\mathbf{R}$-point breathing mode which couples to the charge degree of freedom. We find the same trend in the calculated electron scattering rates due to these phonon modes. These findings help to explain the lower resistivity in DFPT-based transport calculations compared with experiment~\cite{abramovitch_respective_2024, brahlek_hidden_2024}, and highlight the strong effect of electronic correlation in the $e$-ph interactions of multi-orbital correlated metals. 
\\
\indent 
In CCO, we find that correlation within the $d_{x^2-y^2}$ orbital has a modest effect on the coupling of Fermi surface electrons to full- and half-breathing modes but induces a very strong dependence of the EPC on electron frequency. A further surprise is that even though the basic EPC coupling in CCO is Holstein-like (coupling to density) for both breathing modes, which competes with with correlation (acting to suppress density fluctuations), the net effect of the correlations on the EPC at the Fermi energy is a slight suppression for the $\mathbf{M}$-full breathing mode but a slight enhancement for the $\mathbf{X}$-half breathing mode. These differences observed between the two modes suggest that the physics of EPC in the actual material is richer than implied by simple Holstein-Hubbard-type models.
\\
\indent 
A remarkable finding is the significant frequency dependence of the EPC, which leads to significant differences between the value of the near-Fermi surface EPC and the infinite frequency (Hartree) limit and produces variation of the EPC over scales of the order the phonon frequency especially for CCO.  This  strongly frequency-dependent EPC is absent in $+U$ and hybrid extensions of DFT and may have important consequences for physical properties. For example, the associated particle-hole asymmetry in quasiparticle relaxation times could impact properties like thermoelectric transport. For instance, a recent experiment unexpectedly found a positive Seebeck coefficient in a hole-doped cuprate~\cite{gourgout_seebeck_2022}, which the authors explained by suggesting a greater scattering rate above the Fermi energy than below. This would be consistent with our findings of EPC in hole-doped CCO. Other studies in hole-doped cuprates~\cite{doiron-leyraud_hall_2013,cyr-choiniere_anisotropy_2017} have found a transition from negative to positive Seebeck coefficient as temperature increases, which were explained by Fermi surface reconstruction, but could plausibly involve thermal activation of particle-hole asymmetric $e$-ph scattering processes. In this context, our findings that correlation can induce a strong frequency dependence in the EPC motivate further studies on the role of this effect on scattering and transport properties. 
\\
\indent
Taken together, our results highlight the importance of electronic correlation in the $e$-ph physics of correlated metals, which leads to significant renormalization of the couplings at the Fermi surface, a strong electron frequency dependence, and modifications to physical properties. We also demonstrate the dependence of these effects on the microscopic nature of the electronic states and phonon modes considered.  
A strength of our method is the ability to Wannier interpolate the calculated EPC to a complete set of bands and $\mathbf{k}-$points. However, it is limited to a small number of phonon modes and $\mathbf{q}-$points.
This indicates the importance of more efficient computational schemes to calculate EPC at the DMFT level. Furthermore, our findings call for investigation into the effect of different impurity construction and double counting correction schemes and the development of EPC calculations using methods such as charge self-consistent DFT+DMFT or $GW$+DMFT. 

\section*{Acknowledgments}
D.J.A. is supported by the National Science Foundation Graduate Research Fellowship under Grant No. 2139433. 
The Flatiron Institute is a division of the Simons Foundation. 
The work of SB on this project was supported by a grant from the Simons Foundation (00010503, AT).
\\

\bibliography{apssamp}

\begin{thebibliography}{101}
\bibitem [{foo()}]{footnote_on_dfptU}%
  \BibitemOpen
  \href@noop {} {\bibinfo {title} {For consistency with the {DFT+DMFT} values, the {DFPT+$U$} local coupling listed here is calculated with {DFPT} phonon frequencies and eigendisplacements. {For} all of the phonon modes considered, the change in phonon frequency between {DFPT} and {DFPT+$U$} is small (a few {meV} or less).}}\BibitemShut {Stop}%
\end{thebibliography}

\end{document}